\newcommand{\gtrsim}
{\,\raisebox{0.35ex}{$>$}
\hspace{-1.7ex}\raisebox{-0.65ex}{$\sim$}\,
}

\newcommand{\lesssim}
{\,\raisebox{0.35ex}{$<$}
\hspace{-1.7ex}\raisebox{-0.65ex}{$\sim$}\,
}

\documentclass[12pt]{iopart}
\usepackage{graphicx}

\begin{document}

%##########################################################################
\title{Critical dynamics of an interacting magnetic nanoparticle system}

\author{M.~F. Hansen,$^{1,2}$ P.~E. J{\"o}nsson,$^{1}$ P. Nordblad,$^{1}$ and P. Svedlindh$^{1}$}
\address{$^{1}$Department of Materials Science, Uppsala University, Box 534,
SE-751 21 Uppsala, Sweden\\
$^{2}$Department of Physics, Building 307, Technical University of Denmark,
DK-2800 Kgs. Lyngby, Denmark\\
}
\date{\today}
\maketitle
%##########################################################################
%----------------------------------- ABSTRACT
\begin{abstract}
Effects of dipole-dipole interactions on the magnetic relaxation
have been investigated for three Fe-C nanoparticle samples with
volume concentrations of 0.06, 5 and 17~vol\%. While both the 5
and 17~vol\% samples exhibit collective behavior due to dipolar
interactions, only the 17~vol\% sample displays critical behavior
close to its transition temperature. The behaviour of the 5~vol\%
sample can be attributed to a mixture of collective and single
particle dynamics.

\end{abstract}
%##########################################################################
\submitto{\JPCM}

%##########################################################################
%----------------------------------- INTRODUCTION
\section{Introduction}
The dynamics of systems of interacting ferromagnetic nanoparticles
has been the focus of extensive research in recent years. A
question of considerable controversy has been the existence of a
phase transition to a low-temperature spin-glass-like phase \cite{Luo91,MorupEPL94,Dormann97,Djurberg97,Dormann98,JonssonPRL98,MamiyaPRL99,Dormann99,Mamiya99}.
Recent studies on strongly interacting systems have reported a
critical slowing down \cite{Djurberg97,Dormann98,Mamiya99} and a
critical divergence of the non-linear
susceptibility \cite{JonssonPRL98}. 
In addition, non-equilibrium
properties such as ageing, memory and rejuvenation phenomena have
been observed in magnetic relaxation and low frequency AC
susceptibility measurements at low
temperatures \cite{MamiyaPRL99,Joensson00}. These observations
indicate the existence of a low-temperature spin-glass-like state.
However, some questions about the collective state of interacting
nanoparticles are still not resolved. In a recent experimental
study of an interacting maghemite nanoparticle system
\cite{Jonsson98}, it was not possible to find evidence of a finite
temperature transition to a spin-glass-like phase. Still, the
particle system exhibited non-equilibrium dynamics in magnetic
relaxation experiments, typical of spin glasses.  The absence of a
critical behaviour was in this study linked to mixing of collective
and single particle relaxation effects. Mixing of this kind can
for instance originate from a broad particle size distribution, or
it may be observed in a system with a small volume fraction of
particles and therefore comparably weak inter-particle
interactions.

In this work, extensive studies of the magnetic dynamics in a
close to monodisperse nanoparticle system are presented. Three
samples having different volume concentrations of nanoparticles
were investigated; one dilute and nominally non-interacting sample
and two more concentrated interacting samples. Both interacting
samples exhibit collective behaviour due to dipolar interactions at
low temperatures, but only the most interacting sample displays
critical dynamics in the investigated time window and shows equilibrium
dynamics quite similar to that of an atomic spin glass.

%%%%%%%%%%%%%%%%%%%%%%%%%%%%%%%%%%%%%%%%%%%%%%%%%%%%%%%%%%%%%%%%%%%%%%%%%%
%%%%%%%%%%%%%%%%%%%%%%%%%%%%%%%%%%%%%%%%%%%%%%%%%%%%%%%%%%%%%%%%%%%%%%%%%%

\section{Experimental}
The particles were prepared by thermal decomposition of Fe(CO)$_5$
(20.0 ml) in a mixture of carrier liquid (50.0 ml cis-trans
decalin) and surfactant (4.0 g oleic acid) by the method described
in \cite{Wonterghem88}. This method leads to the
formation of surfactant coated particles of the amorphous alloy
Fe$_x$C$_{1-x}$ (x $\approx$ 0.2-0.3). The decomposition was
carried out slowly by using low heating powers. Great care was
taken at all times to avoid oxidation of the particles and a flow
of Ar/H$_2$ (98/2\%) gas through the reaction vessel was
maintained throughout the preparation. After preparation, the
decalin carrier liquid was evaporated at 140 $^\circ$C under
reduced pressure using a gas flow to drive out the vapour and a
small amount of de-gassed xylene was added such that the volume
fraction of particles in the resulting base ferrofluid was about 5
vol\%. More dilute samples were obtained by further dilution of
this ferrofluid by addition of xylene and more concentrated
samples were obtained by evaporating the xylene at room
temperature in vacuum. All sample handling after the preparation
of the base ferrofluid was carried out in an argon glove box. The
chemical state of the iron in the ferrofluids was checked by
M{\"o}ssbauer spectroscopy, which showed that no significant
oxidation of the particles had taken place during the sample
preparation and sample handling. The particle volume fractions in
the resulting samples were estimated from the iron content
determined by atomic absorption spectroscopy. The sizes of the
particles were determined from transmission electron microscopy
(TEM) studies. A droplet of a dilute ferrofluid was placed onto a
grid, that was left in air for about a week to ensure full
oxidation of the particles. The size determination from the
resulting TEM micrographs was subsequently corrected for the
change in density due to the oxidation of the particles. The shape
of the particles was spherical to a good approximation. The
particle size distribution was obtained using the computer
analysis method described in \cite{FiskerJNR00}. The
analysis of 1579 particles yielded the average particle diameter
$d =5.3$ nm with the standard deviation 0.3 nm. The
volume-weighted volume distribution was well described by the
lognormal distribution, $f(V)dV = (2\pi)^{-1/2}(\sigma
V)^{-1}\exp[-\ln^2(V/V_m)/2\sigma^2]dV$, with $V_m = 78.2$ nm$^3$
and $\sigma = 0.13$. The saturation magnetisation was estimated to
$1\times 10^6$ Am$^{-1}$ from Langevin fits to magnetisation vs.
field curves at several temperatures well above the
superparamagnetic blocking temperature. The present preparation
batch is slightly different from that studied previously in
\cite{Djurberg97,JonssonPRL98,HansenJMMM98}. We have
studied three samples with particle concentrations of $17\pm4$,
$5\pm1$ and $0.06\pm0.02$~vol\%. In the most concentrated sample,
most of the carrier liquid was evaporated and the sample had a
paste-like consistency. This resulted in poor thermal contact
between the sample and the sample container. Thermal stability was
not achieved until several minutes after a change in temperature,
preventing systematic studies of the non-equilibrium dynamics for
this sample.

Two experimental set-ups were used for the magnetic studies. A
commercial susceptometer (LakeShore model 7225) was used for DC field
scans and AC susceptibility measurements in a frequency range of 15 -
1000 Hz. A non-commercial superconducting quantum interference device
(SQUID) magnetometer \cite{Magnusson97} was used for AC susceptibility
measurements in a frequency range of 10 mHz - 9.1 kHz, zero field cooled
(ZFC) relaxation measurements and magnetic noise measurements. The
applied AC and DC fields were chosen small enough (0.1 Oe) to ensure linear
response from the samples. The background field was less than 1 mOe. The ZFC
relaxation measurements were carried out by cooling the sample to the
measuring temperature $T$ in zero field, equilibrating the system
for a wait time $t_w$, and then applying a small DC field $h$ and
measuring the magnetisation $M(T,t)$ as a function of time, $t$, after
that the field was applied. For a slowly relaxing system, it can be
shown that $M(T,t)$ and the relaxation rate $S(T,t) \equiv h^{-1}
\partial M(T,t)/\partial\ln t$ relate to the real and imaginary
components of the AC susceptibility as \cite{Lundgren81}
%%%%%%%%%%%%%%%%%%%%%%%%%%%%%%%%%%%%%%%%%%%%%%%%%%%%%%%%%%%%%%%%%%%%%%
\begin{eqnarray}
M(T,t)/h &\approx& \chi '(T,\omega),\\
S(T,t) &\approx&{2\over\pi}\chi''(T,\omega),
\label{S}
\end{eqnarray}
%%%%%%%%%%%%%%%%%%%%%%%%%%%%%%%%%%%%%%%%%%%%%%%%%%%%%%%%%%%%%%%%%%%%%%
with $t = 1/\omega$.
Magnetic noise measurements were performed in zero external field
using an HP35670A dynamic signal analyser. The power spectrum of the
magnetic fluctuations was measured in three overlapping frequency
intervals (i) 0.003-12.5 Hz, (ii) 0.25-400 Hz and (iii) 8-12500 Hz.
The background spectra were subtracted from the data.
Data of the same order of magnitude as the
background signal were not used in the analysis. The
fluctuation-dissipation theorem \cite{Callen51} relates the noise
power spectrum to the zero-field limit of the out-of-phase AC
susceptibility as
%%%%%%%%%%%%%%%%%%%%%%%%%%%%%%%%%%%%%%%%%%%%%%%%%%%%%%%%%%%%%%%%%%%%%%
\begin{equation}
P(T,\omega)=4k_{B}T{{\chi''(T,\omega)}\over{\omega}}.\label{eq_noise}
\end{equation}
%%%%%%%%%%%%%%%%%%%%%%%%%%%%%%%%%%%%%%%%%%%%%%%%%%%%%%%%%%%%%%%%%%%%%%
A comparison between the out-of-phase component obtained from AC
susceptibility measurements and results obtained from zero field
noise measurements verifies that equation (\ref{eq_noise}) is obeyed
and hence that the AC data are obtained in the linear response
regime.
%%%%%%%%%%%%%%%%%%%%%%%%%%%%%%%%%%%%%%%%%%%%%%%%%%%%%%%%%%%%%%%%%%%%%
%%%%%%%%%%%%%%%%%%%%%%%%%%%%%%%%%%%%%%%%%%%%%%%%%%%%%%%%%%%%%%%%%%%%%

\section{Results and discussion}

\subsection{General behaviour}
\label{GeneralBehavior} 
\begin{figure}[htb]
\begin{center}
\includegraphics[width=0.8\textwidth]{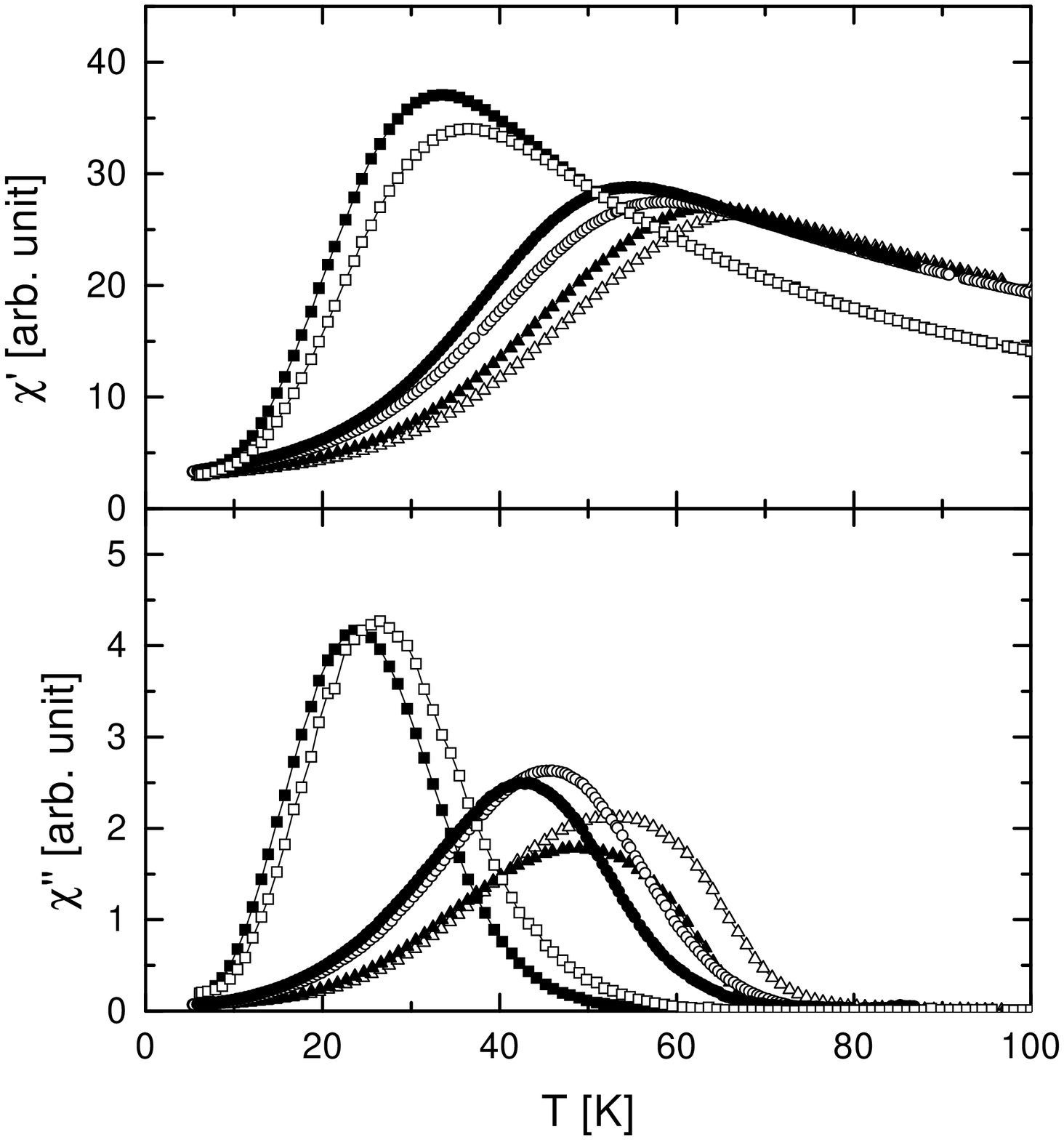}
\caption[]{AC susceptibilities for the 0.06~vol\% ($\opensquare$),
5~vol\% ($\circ$), and 17~vol\% ($\triangle$) samples at frequencies of
$f = $125 Hz (filled symbols) and $f = 1000 $Hz (open symbols).}
\label{fig_x3_ac}
\end{center}
\end{figure}

Figure \ref{fig_x3_ac} shows the AC
susceptibility measured in the commercial set-up for the three
samples. The most dilute sample with a 0.06~vol\% fraction of
particles is intended to serve as an experimental reference for a
non-interacting system. However, the peak height of $\chi''$ for
this sample increases slightly with increasing frequency. This
trend is opposite to the expected behaviour for non-interacting
nanoparticles \cite{raiste97} and thus indicates that the dynamics
of this sample is influenced by weak inter-particle interactions.
Despite this, it is possible to assess the distribution of energy
barriers with reasonable accuracy making use of the method
described in \cite{JonssonJMMM97}. This analysis yields
$\tau_0 = 1\times 10^{-12}$ s and $K = 0.9\times 10^5$ Jm$^{-3}$.
The extracted values of $K$ and $\tau_0$ compare reasonably well with
previous estimates. For 5.0 nm Fe-C particles, $K$ has been
estimated to $1.3\times 10^5$ Jm$^{-3}$, and it has been found
that $K$ increases with decreasing particle size to $3\times10^5$
Jm$^{-3}$ for 3.2 nm particles \cite{HansonJPC93,HansonJPC95}.
Studies of different preparation batches have reported values of
$\tau_0$ in the range $2\times 10^{-12}$ s to $3\times 10^{-11}$
s \cite{Djurberg97,HansenJMMM98,HansonJPC95,LinderothJMMM93}.

\begin{figure}[htb]
\includegraphics[width=0.8\textwidth]{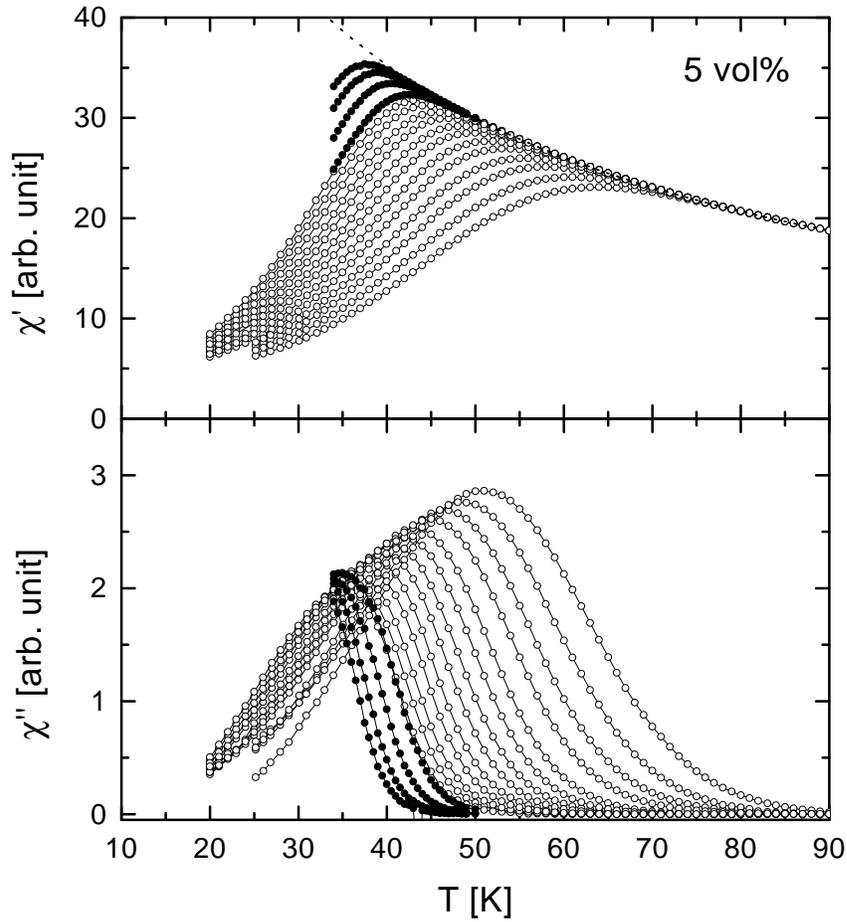}
\caption{AC susceptibilities for the 5~vol\% sample using the frequencies
(left to right, open symbols) $f = $0.010, 0.031, 0.091, 0.31,
0.91, 3.1, 9.1, 31, 91, 310, 910, 3100, 9100 Hz. The filled points
are obtained from ZFC relaxation measurements and correspond to
the frequencies $f = $0.019, 0.103, 0.979, 10.1 mHz. The dashed
line indicates the equilibrium susceptibility.} \label{fig_x1_ac}
\end{figure}

\begin{figure}[htb]
\includegraphics[width=0.8\textwidth]{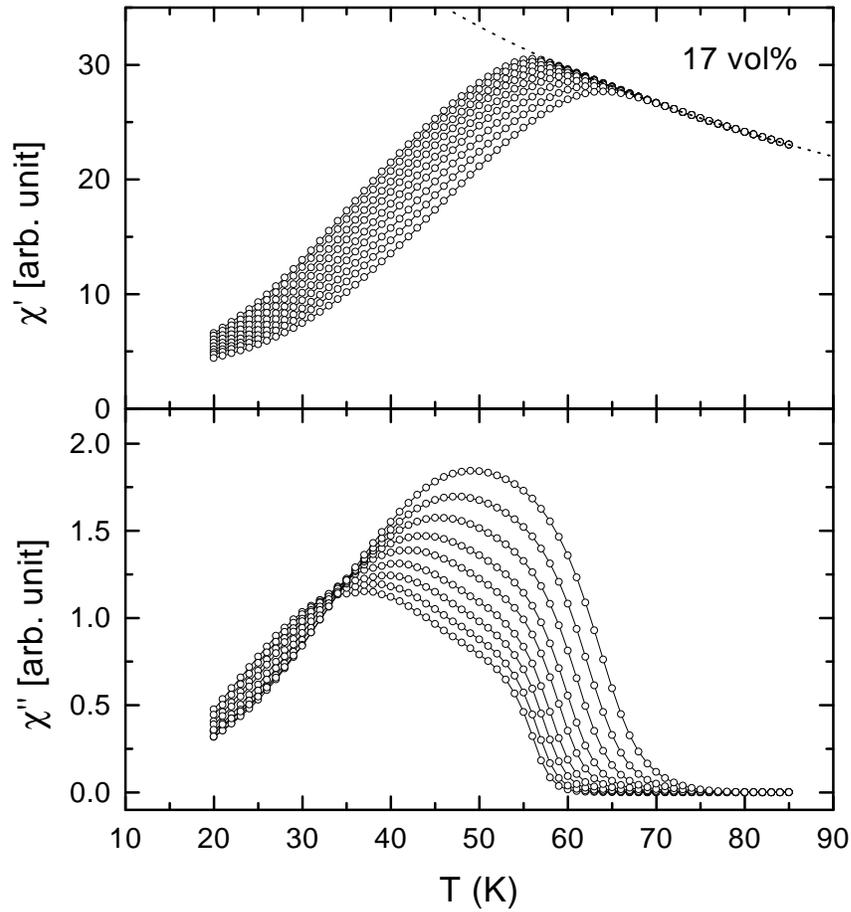}
\caption{AC susceptibilities for the 17~vol\% sample using the frequencies
(left to right) $f = $0.017, 0.051, 0.17, 0.51, 1.7, 5.1, 17, 51,
170 Hz. The dashed line indicates the equilibrium susceptibility.}
\label{fig_dry_ac}
\end{figure}

Figures \ref{fig_x1_ac} and \ref{fig_dry_ac} show the AC
susceptibility data measured in the non-commercial SQUID for the 5
and 17~vol\% samples, respectively. For the 5~vol\% sample, ZFC
relaxation data are also included.  A few differences in the
behaviour of these samples as compared to the dilute sample can
immediately be noticed. The peak height of $\chi ''$ increases
significantly with increasing frequency, and the width of the peak
increases with increasing frequency. The onset of a non-zero
$\chi''$, is shifted towards higher temperatures and it becomes
sharper with increasing volume fraction of particles.

\begin{figure}[htb]
\includegraphics[width=0.8\textwidth]{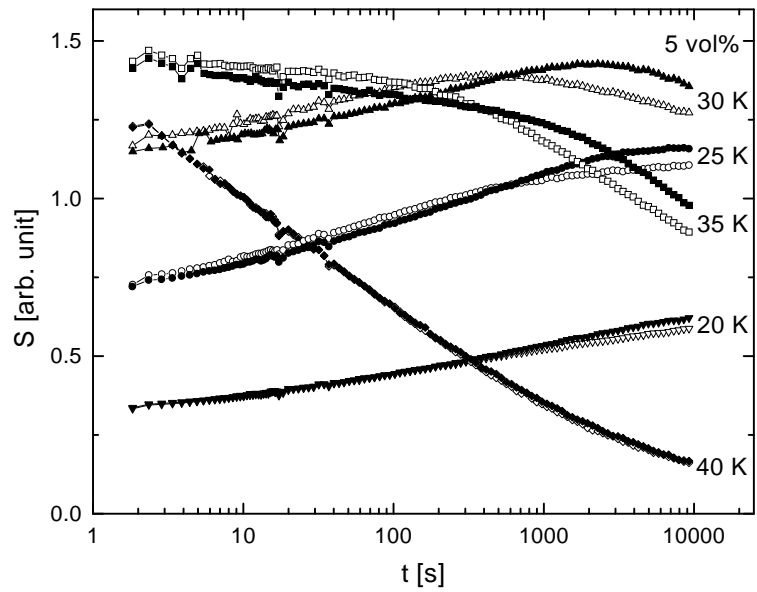}
\caption{Relaxation rate for the 5~vol\% sample obtained from ZFC
relaxation measurements at the indicated temperatures after wait
times of 300 s (open symbols) and 3000 s (filled symbols).}
\label{fig_x1_age}
\end{figure}

Figure \ref{fig_x1_age} shows the relaxation rate for the 5~vol\%
sample measured for two different wait times at different
temperatures. Below 40 K, a clear wait time dependence is observed
indicating that non-equilibrium phenomena plays a key role for the
dynamics at low temperatures. The same sample has recently been
subject to a detailed study of the low temperature non-equilibrium
dynamics in which the memory effect in the ac susceptibility was
observed and further characterised by temperature cycling ZFC
relaxation experiments \cite{Joensson00}. It was shown that the
non-equilibrium dynamics in this sample is governed by long-range
collective behaviour and that well-known concepts for spin glasses
such as chaos with temperature and overlap length are necessary to
describe the observations. This implies that the magnetic
relaxation of the two concentrated samples must be analysed in
terms of collective dynamics. A model for weakly interacting
particles, such as that proposed in
\cite{JonssonEPL01}, where the effect of dipolar interaction
on the relaxation time is accounted for by introducing the
thermodynamic averages of the local dipolar field in a rigorous
expression for the single particle relaxation time, is only valid
at temperatures much higher than the freezing temperatures of the
two concentrated samples. Below, we briefly introduce the concepts
of dynamic scaling and discuss the observed dynamics for the two
concentrated samples in terms of critical dynamics.

%%%%%%%%%%%%%%%%%%%%%%%%%%%%%%%%%%%%%%%%%%%%%%%%%%%%%%%%%%%%%%%%%%%%%%
%%%%%%%%%%%%%%%%%%%%%%%%%%%%%%%%%%%%%%%%%%%%%%%%%%%%%%%%%%%%%%%%%%%%%%

\subsection{Dynamic scaling}
\label{DynScaling} A signature of a continuous magnetic phase
transition is the divergence of the correlation length, $\xi$,
when the phase transition temperature, $T_g$, is approached from
above as $\xi/a = \epsilon^{-\nu}$, where $a$ is the average
distance between interacting moments, $\epsilon = T/T_g-1$ is the
reduced temperature and $\nu$ is a critical exponent. According to
conventional critical slowing down, the longest relaxation time
due to correlated dynamics, $\tau_c$, is related to the
correlation length as $\tau_c \propto (\xi/a)^z$, where $z$ is the
dynamic critical exponent. Hence, for $T\rightarrow T_g^+$
%%%%%%%%%%%%%%%%%%%%%%%%%%%%%%%%%%%%%%%%%%%%%%%%%%%%%%%%%%%%%%%%%%%%%%
\begin{equation}
\tau_c = \tau_* \epsilon^{-z\nu},
\label{eq_ds}
\end{equation}
%%%%%%%%%%%%%%%%%%%%%%%%%%%%%%%%%%%%%%%%%%%%%%%%%%%%%%%%%%%%%%%%%%%%%%
where $\tau_*$ is a microscopic relaxation time. According to the dynamic scaling
hypothesis \cite{Hohenberg77}, for $T\rightarrow T_g^+$ and
$t/\tau_* \gg 1$, the spin auto-correlation function can be
written in the scaling form \cite{Ogielski85}
%%%%%%%%%%%%%%%%%%%%%%%%%%%%%%%%%%%%%%%%%%%%%%%%%%%%%%%%%%%%%%%%%%%%%%
\begin{equation}
q(t)=t^{-\beta/z\nu}Q(t/\tau_c), \label{eq_qt}
\end{equation}
%%%%%%%%%%%%%%%%%%%%%%%%%%%%%%%%%%%%%%%%%%%%%%%%%%%%%%%%%%%%%%%%%%%%%%
where $\beta$ is a critical exponent and $Q(x)$ is a scaling
function. Using linear response theory it is possible from this
relation to obtain the complex susceptibility and derive the
scaling relation \cite{Rigaux95}
%%%%%%%%%%%%%%%%%%%%%%%%%%%%%%%%%%%%%%%%%%%%%%%%%%%%%%%%%%%%%%%%%%%%%%
\begin{equation}
{{\chi''(T,\omega)}\over{\chi_{\rm eq}(T)}}=\epsilon^\beta
G(\omega\tau_c),\label{eq_fd}
\end{equation}
%%%%%%%%%%%%%%%%%%%%%%%%%%%%%%%%%%%%%%%%%%%%%%%%%%%%%%%%%%%%%%%%%%%%%%
where $\omega=1/t$ and $G(x)$ is a scaling function. The
asymptotic behaviour is $G(x) \propto x^y$, with $y=1$ and
$y=\beta/z\nu$ for small and large values of $x$, respectively.
Using the asymptotic behaviour of $G(x)$ in the limit $\omega\tau_c
\rightarrow 0$, the following relation holds
\cite{Ogielski85,Rigaux95,Gunnarsson88}
%%%%%%%%%%%%%%%%%%%%%%%%%%%%%%%%%%%%%%%%%%%%%%%%%%%%%%%%%%%%%%%%%%%%%%
\begin{equation}
{1\over\omega}{{\chi''(T,\omega)}\over{\chi_{\rm eq}(T)}}\propto
{\epsilon^{-z\nu+\beta}\propto{\tau_c^{1-\beta/z\nu}}},
\label{eq_ad}
\end{equation}
%%%%%%%%%%%%%%%%%%%%%%%%%%%%%%%%%%%%%%%%%%%%%%%%%%%%%%%%%%%%%%%%%%%%%%
implying that the left hand side of equation (\ref{eq_ad}) at each
temperature reaches a frequency independent plateau.

The meaning of $\tau_*$ is that it is the relaxation time of the
individual magnetic entities in the system. For spin glasses,
$\tau_* \sim 10^{-13}$ s and is the fluctuation time of an atomic
moment. For nanoparticles, $\tau_*$ can be assigned to the
superparamagnetic relaxation time of a single particle of average
size. In a dense system, the dipolar interaction may modify this
relaxation time compared that of isolated particles. However, as a
first approximation, it is reasonable to assume that $\tau_*$ is
close to the superparamagnetic relaxation time of an isolated
particle, which in the relevant temperature range for our studies
can be approximated by the Arrhenius-Ne{\'el} expression. Below,
we compare two approximations: i) $\tau_* =$ constant, and ii)
$\tau_* = \tau_0 \exp(KV_{m}/k_B T)$. The first approximation has
been used in previous
work \cite{Djurberg97,Dormann98,JonssonPRL98,Mamiya99,JoenssonICM00},
but it is only a good approximation if there is little variation
of the single particle relaxation time in the temperature interval
used for the analysis.

\begin{figure}[htb]
\includegraphics[width=0.8\textwidth]{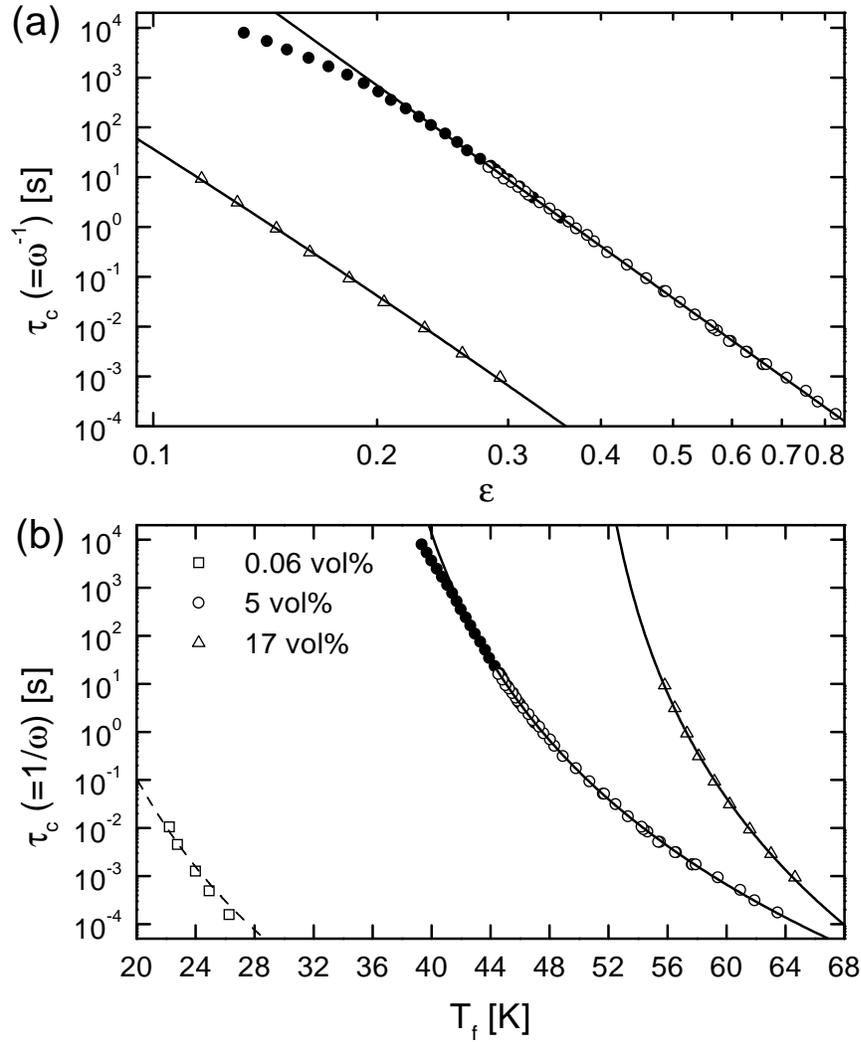}
\caption{a) Relaxation time $\tau_c = \omega^{-1}$ vs. reduced
temperature for the 5  and 17~vol\% samples. The lines are fits to
equation (\ref{eq_ds}) as described in the text. b) Relaxation time vs
temperature for the 0.06, 5 and 17~vol\% samples. The freezing
temperatures for the 5 and 17~vol\% samples were obtained from
$\chi'$ as described in the text. For the 0.06~vol\% sample, the
data correspond to the peak temperatures of $\chi''$, and the
Arrhenius-N{\'e}el expression with parameter values as given in
the text is shown as a dashed line. } \label{fig_ds}

\end{figure}

In AC susceptibility experiments, the slowing down of the
relaxation time $\tau_c$ (equation (\ref{eq_ds}) can be derived from
the temperatures where an onset of dissipation occurs (freezing
temperatures) as a function of the observation time $\omega^{-1}$.
We have considered two criteria for the onset. In the first
criterion, the freezing temperature is defined as the temperature
at which $\chi''(T,\omega)$ attains 15\% of its maximum value. In
the second criterion, the freezing temperature is defined from the
relation $\chi'(T_f,\omega)=0.98\chi_{eq}(T_f)$.
Figure \ref{fig_ds}(b) shows the freezing temperatures for the two
concentrated samples obtained from $\chi'$ data. The
superparamagnetic blocking temperatures estimated from the peaks
of the $\chi''$ data for the dilute sample are included for
comparison.

\subsubsection{17~vol\% sample}
First, we consider the approximation $\tau_* =$ constant. Using
the out-of-phase susceptibility data, a dynamic scaling analysis
according to equation (\ref{eq_ds}) results in $T_g = 49.5 \pm 2$ K,
$z\nu = 10.5 \pm 2$ and $\tau_* = 10^{-7.7 \pm 1}$ s
\cite{JoenssonICM00}. This analysis is performed for reduced
temperatures in the range $0.16 \lesssim \epsilon \lesssim 0.37$
and corresponds to 4 decades of observation times. A similar
analysis using the in-phase susceptibility data results in $T_g =
50.5 \pm 2$~K, $z\nu = 9.5 \pm 2$ and $\tau_* = 10^{-8.3 \pm 1}$
s. The values of $T_g$ and $z\nu$ from the two analyses are in
good agreement, and the value of $\tau_*$ is, considering the
temperature interval used in the analysis and the estimates of
$\tau_0$ and $K$ given above, within the limits implied by the
Arrhenius-N{\'e}el expression. The derived values of $z\nu$ also
compare well to values found in previous work on nanoparticles:
$z\nu = 11 \pm 3$ in \cite{Djurberg97} and $z\nu = 10.5
\pm 3$ in \cite{Mamiya99}, but less well to $z\nu = 7.0
\pm 0.3$ found in reference \cite{Dormann98}. A good agreement is
also found comparing with canonical 3D Ising and Heisenberg spin
glasses, for which $z\nu = 8-10$ \cite{Rigaux95,Gunnarsson88}. 
An analysis according to the full scaling relation,
equation (\ref{eq_fd}), using $T_g = 49.5 \pm 1.5$ and $z\nu = 10.5\pm
2.0$ yields data collapse for $\beta = 1.1 \pm 0.2$ (see reference
\cite{JoenssonICM00}, figure 3). This analysis is based on all
available $\chi''$ data with temperatures corresponding to
$\epsilon > 0.01$ ($T > 50$ K). The value of $\beta$ is in good
agreement with $\beta = 1.2 \pm 0.1$ obtained from a full scaling
analysis of the non-linear magnetic susceptibility on a similar
sample \cite{JonssonPRL98}, but larger than typical values of
$\beta=0.5-0.8$ reported for 3D Ising and Heisenberg spin glasses
\cite{Rigaux95,Geschwind90,Gunnarsson91}. The value of $\beta/z\nu \approx
0.11$, extracted from the asymptotic behaviour of $G(x)$ for large
$x$, is consistent with the derived values of $\beta$ and $z\nu$.

\begin{figure}[htb]
\includegraphics[width=0.8\textwidth]{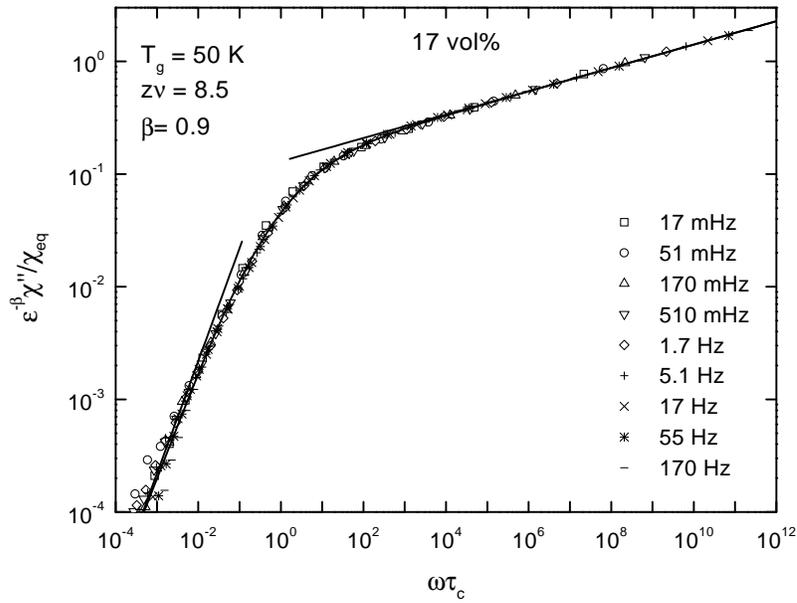}
\caption{Full dynamic scaling plot for the 17~vol\% sample
according to equation (\ref{eq_fd}) using the Arrhenius-N{\'e}el
expression for $\tau_*$ as described in the text. The asymptotic
behaviours of $G(x)$ are shown as straight lines.}
\label{fig_dry_fd}
\end{figure}

Next, we consider the effect of using $\tau_* =
\tau_0\exp(KV_{m}/k_{B}T)$. It should be noted that this
introduces one extra parameters that can be varied in the
analysis, and we have therefore explored the critical behaviour for
a variety of choices of $\tau_0$ and $KV_{m}$ to evaluate the
robustness of the estimates of $z\nu$ and $\beta$. Critical
slowing down analyses have been performed for various fixed values
of $\tau_0$ and $T_g$ and the values of $z\nu$ and $KV_{m}$ have
been estimated and used in the analysis according to the scaling
relation, equation (\ref{eq_fd}), to extract the value of $\beta$. From
these analyses, it is found that both the critical slowing down
analysis and the full scaling relation can be fulfilled for $T_g =
50 \pm 2$ K with $\tau_0=1\times10^{-11} s$. The estimate of
$KV_{m}/k_{B}$ depends slightly on whether the $\chi'$ or the
$\chi''$ data are used in the critical slowing down analysis, and
the corresponding parameter intervals are $KV_{m}/k_{B}=500\pm
100$ K using $\chi'$ (shown in figure \ref{fig_ds}) and
$KV_{m}/k_{B}=650\pm 100$ K using $\chi''$. For both analyses, the
extracted critical exponents attain the values $z\nu = 8.5 \pm 2$
and $\beta = 0.9 \pm 0.2$. Other values of $\tau_0$ of the order
of $10^{-11}$ s yield slightly different values of $KV_m/k_B$ but
the same values of the critical exponents. The values of
$KV_{m}/k_{B}$ and $\tau_0$ are of the same size as the estimates
for the dilute sample, $KV_{m}/k_{B} \approx 510$ K and $\tau_0
\sim 1\times10^{-12}$ s, but it is stressed that an exact
correspondence is not expected. The values of $z\nu$ and $\beta$
are slightly smaller than those obtained from the analysis
assuming a constant $\tau_*$. The decrease of $z\nu$ is due to the
fact that the temperature dependence of $\tau_c/\tau_*$ is weaker
when $\tau_*$ is allowed to vary with temperature. The reduced
value of $z\nu$ and the extracted smaller value of $\beta$ leads
to $\beta/z\nu\approx 0.11$, which is the same value as found from
the analysis using a constant value of $\tau_*$. The data collapse
of $\epsilon^{-\beta}\chi''(T)/\chi_{eq}(T)$ vs. $\omega\tau_c$ to
a single function $G(x)$ according to equations (\ref{eq_ds}) and
(\ref{eq_fd}) using $\tau_0 = 10^{-11}$ s and $KV_{m}/k_{B} = 570$
K (the average of the values obtained using $\chi'$ and $\chi''$
in the critical slowing down analyses) is shown in figure
\ref{fig_dry_fd}. The scaling is of the same quality as that
assuming a constant $\tau_*$, and the asymptotic behaviours are the
same (in agreement with the estimates of $\beta$ and $z\nu$).

\begin{figure}[htb]
\includegraphics[width=0.8\textwidth]{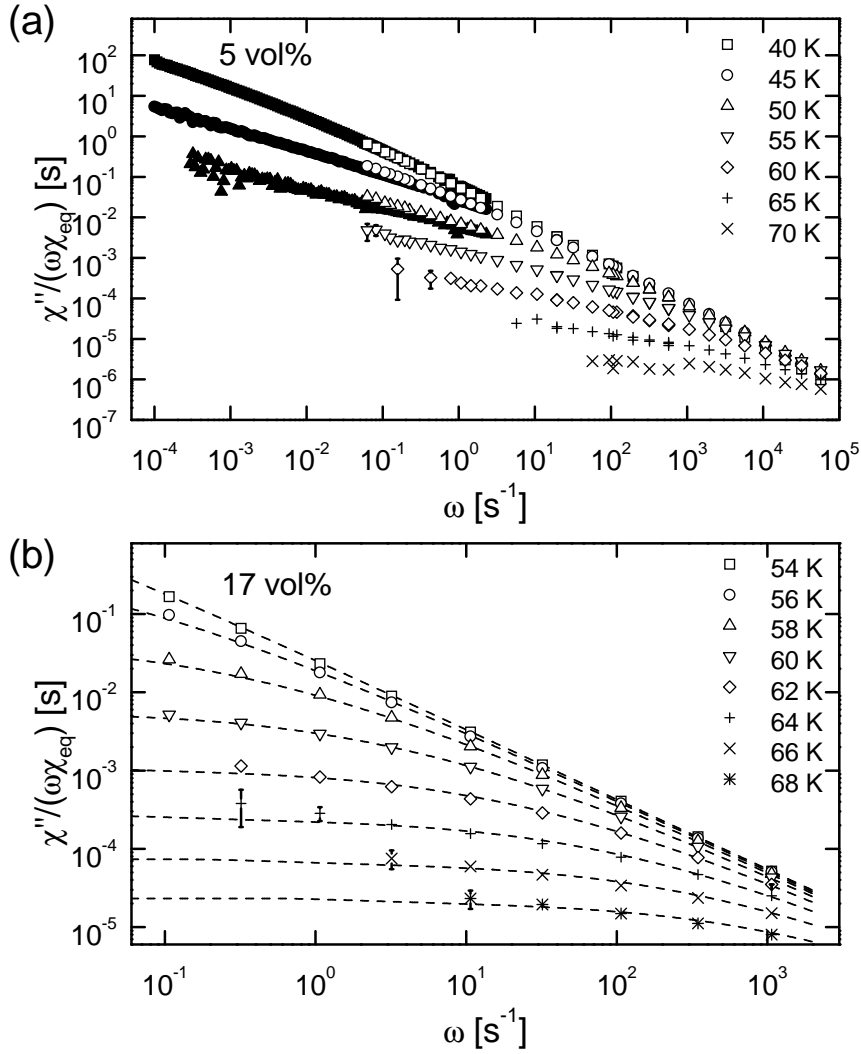}
\caption{(a) Plot of the AC data (open points) and ZFC relaxation
data (filled points) according to equation (\ref{eq_ad}) for the 5
vol\% sample. (b) Plot of the AC data according to
equation (\ref{eq_ad}) for the 17~vol\% sample. The dashed lines are
obtained from $G(x)$ with the expected asymptotic behaviours.
Errorbars are only shown when larger than the symbol size.}
\label{fig_ad}
\end{figure}

Figure \ref{fig_ad}(b) shows $\chi''(\omega)/\omega\chi_{eq}$ as a
function of $\omega$ for different temperatures $T>T_g$ in a
log-log plot for the 17~vol\% sample. The prediction of plateaus
at low frequencies from equation (\ref{eq_ad}) holds regardless of a
temperature dependence of $\tau_*$ and figure \ref{fig_ad}(b) shows
that the prediction is well confirmed for this sample. In the same
figure, the behaviour calculated from
equation (\ref{eq_fd}) is also included using the experimentally determined scaling
function, $G(x)$, and extrapolating using its asymptotic behaviour.

\subsubsection{5~vol\% sample}

For the 5~vol\% sample, an apparent scaling according to
equation (\ref{eq_ds}) with a constant value of $\tau_*$ can be
obtained for temperatures corresponding to $0.25 \lesssim \epsilon
\lesssim 0.8$ (6 decades of observation times) with $T_g = 35.1
\pm 2$ K, $z\nu = 10.8 \pm 1$ and $\tau_* = 10^{-4.5 \pm 0.5}$ s
using $\chi''$ data, and with $T_g = 34.6 \pm 2$~K, $z\nu = 10.8
\pm 1$ and $\tau_* = 10^{-4.7 \pm 0.5}$ s using $\chi'$ and ZFC
relaxation data (shown in figure \ref{fig_ds}). However, deviations
from scaling are found at lower temperatures ($\epsilon < 0.25$).
Moreover, it is not possible to obtain data collapse
according to the full scaling relation equation (\ref{eq_fd}), with
temperatures corresponding to $0.02 \lesssim \epsilon \lesssim
0.8$ (or to $0.25 \lesssim \epsilon \lesssim 0.8$), for any choice
of $T_g$, $z\nu$ and $\beta$. Based on the estimates of $KV_m$ and
$\tau_0$ for the dilute sample and the temperature range used in
the analysis above, it is expected that $\tau_* \sim
10^{-9} - 10^{-7}$ s. However, the derived value ($\tau_* \sim
10^{-4.6}$) deviate by
orders of magnitude from this range. Furthermore, even introducing an
Arrhenius-N{\'e}el temperature dependence of $\tau_*$, it is not
possible to obtain unambiguous parameters from the critical slowing
down analysis and to fulfil scaling according to
equation (\ref{eq_fd}). A
simple manifestation of critical dynamics, which is independent of
a temperature dependence of $\tau_*$, is that
$\chi''(\omega)/\omega\chi_{eq}$ should settle on plateaus for
small values of $\omega$. Conferring figure \ref{fig_ad}(a), it is
seen that this prediction is not fulfilled for the 5 vol\% sample
(except for $T \geq 65$ K), and this is a further indication of
non-critical dynamics in this sample.

Several factors may contribute to deviations from critical
behaviour and a possible non-divergence of the correlation length.
First, there may be physical clustering of the particles ({\it
i.e.}, regions with a higher density of particles than average).
The effect of clustering is two-fold: At comparably high
temperatures, where short-range correlations are relevant, the
stronger inter-particle interaction in particle clusters will
enhance the local correlation length and shift the dynamics to
longer time scales than expected from the volume fraction of
particles and a homogeneous particle dispersion. In addition, the
small scale heterogeneity will create a dispersion of length
scales for the collective dynamics that modifies and possibly
limits the growth of correlations at lower temperatures. Second,
the polydispersivity of the particle system leads to a
distribution of single particle relaxation times with a width that
increases significantly with decreasing temperature (and is thus
more important for the 5 vol\% sample than for the 17 vol\%
sample). At lower temperatures, the largest particles may
therefore become thermally blocked on time scales comparable to
the longest time scale related to the collective dynamics and act
as random magnets instead of taking part in the collective
dynamics. This may completely mask or even obstruct long-range
collective behaviour, as discussed in \cite{Jonsson98}.
The non-critical behaviour of the present 5 vol\% sample is most
likely both due to the formation of clusters of particles (see
section below) and the dispersion of single particle relaxation
times.

\subsection{Relaxation function}

For a spin glass and $T \lesssim T_{g}$, $S(T,t)$ is history
dependent and has a non-trivial variation with $t$. However, when
quasi-equilibrium is probed, one can write $S(T,t) \propto
t^{-y(T)}$, where $y(T)$ attains a positive value close to
zero \cite{Ogielski85}. For $T > T_{g}$, the time dependence
of $S(T,t)$ is determined by the critical dynamics. As discussed
previously, the critical dynamics result in fluctuations spanning
all time scales between $\tau_*$ and $\tau_c$, and this gives a
very broad spectrum of relaxation times. From equation (\ref{S}) and
using the asymptotic behaviours of $G(x)$ in equation (\ref{eq_fd}), it
is easily seen that $S(T,t) \propto t^{-y(T)}$ with $y(T) =
\beta/z\nu \approx 0.1$ for $\tau_* \ll t\ll \tau_c$ and $y(T) =
1$ for $t\gg\tau_c$ \cite{Note1}. 
At intermediate observation
times, $y(T)$ monotonically increases with $t$ between the two
extremes and a cross-over (or 'knee') from an essentially flat
$S(T,t)$ vs. $t$ curve to $S(T,t)\propto t^{-1}$ is expected at $t
\sim \tau_c$.

\begin{figure}[htb]
\includegraphics[width=0.8\textwidth]{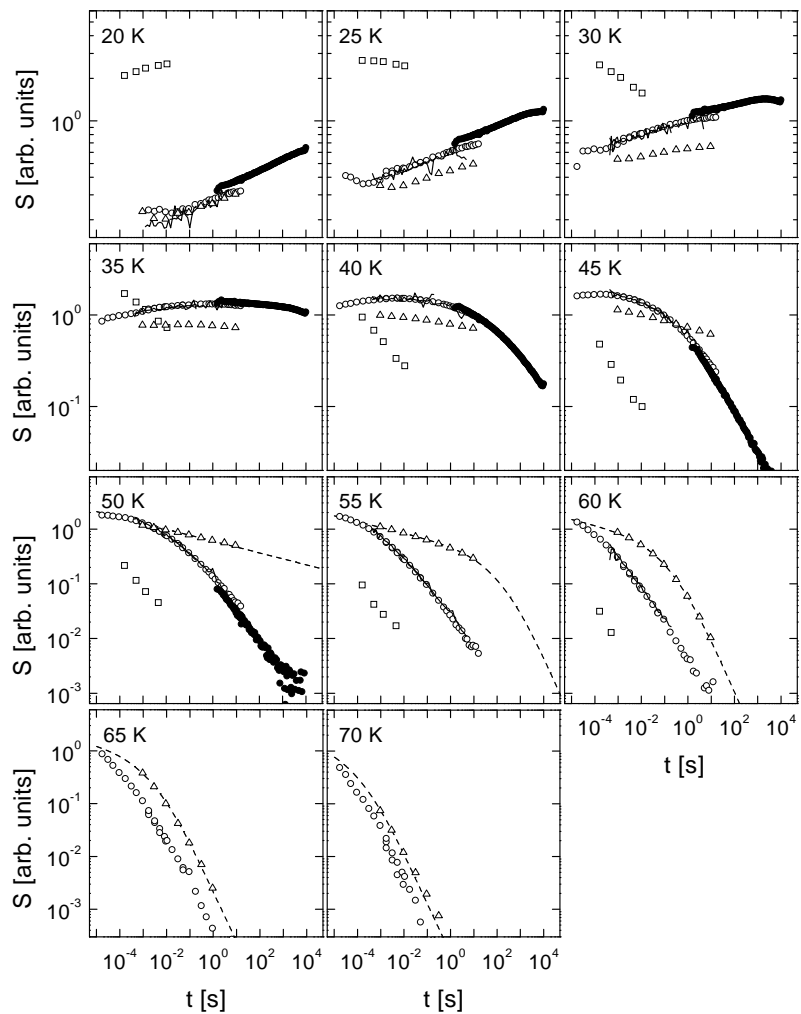}
\caption{Relaxation rate vs. observation time at different
temperatures. For the 0.06~vol\% sample ($\opensquare$) the relaxation
rate is calculated from the AC susceptibility data. For the 5
vol\% sample ($\circ$), the relaxation rate is calculated from AC
susceptibility data (open symbols), ZFC data (filled symbols) and
noise data (lines)  and for the 17~vol\% sample the relaxation
rate is calculated from AC susceptibility data ($\triangle$) and
from $G(x)$ with the expected asymptotic behaviours (dashed lines).} 
\label{fig_com_s}
\end{figure}

Figure \ref{fig_com_s} shows the relaxation rate vs time for the
three samples at different temperatures in the range $20$~K $\leq
T \leq$ $70$~K. For the 0.06 and 17~vol\% samples $\chi''$ data
are shown, and for the 5~vol\% sample results from ZFC relaxation,
AC susceptibility and magnetic noise measurements are included.
For the 17~vol\% sample,  $\chi''$ data for $T>T_g$ obtained from
equation (\ref{eq_fd}) using the experimentally estimated $G(x)$ and
the expected asymptotic behaviour for small $x$ are also shown. In
the low-temperature region, $T \lesssim 30$ K, the magnitudes of
the relaxation rates for the two interacting samples are much
smaller than that for the 0.06~vol\% sample. This is a signature
of collective dynamics \cite{Jonsson98,Andersson97}. Moreover, the
relaxation rate of the 17~vol\% sample is smaller than that of the
5~vol\% sample. For temperatures $T \gtrsim 40$~K and in the
investigated time window, the relaxation rate of the 0.06~vol\%
sample decays with increasing observation time, while the
relaxation rate of the the 5~vol\% sample at short time scales
exhibits a weak frequency dependence followed by a knee and a
decrease towards zero at longer time scales. This approach towards
zero relaxation rate ($y(T) \approx 0.5-0.6$ between 45 and 55 K)
is slower than that of an atomic spin glass (for which $y(T)=1$)
and consistent with the lack of plateaus in the
$\chi''(\omega)/\omega\chi_{eq}$ curves for the corresponding
temperatures in figure \ref{fig_ad}(a). The time dependence of the
relaxation rate at time scales shorter than this knee is also
uncharacteristic of a spin glass as it is non-monotonic in time
and shows a broad maximum. For the 17~vol\% sample and in the
temperature range $35$~K $\lesssim T \lesssim 50$~K, the
relaxation rate exhibits a slow, monotonous decrease with
increasing observation time. The relaxation rate follows a $S(T,t)
\propto t^{-y(T)}$ dependence, with $y(T)$ decreasing with
decreasing temperature, mimicking the expected behaviour of a spin
glass system \cite{Ogielski85}. 
It should be noted though, that the
particle size distribution will cause a temperature dependent and
broad onset of the response on short observation times, as can be
envisaged by the relaxation rate curves at the lowest
temperatures. At temperatures above $T_g$, $T > 50$ K, a knee
appears in the relaxation rate also for this sample and the
approach towards zero follows a $S(T,t) \propto t^{-1}$. At
temperatures $T \geq 65$ K, the cut-off in the relaxation rate is
equally sharp for both interacting samples, and the dynamics of
the two samples are rather similar. This indicates, for the 5
vol\% sample, that the particle moments participating in the
dynamics on long time scales in this temperature range are more
strongly interacting than average, {\it i.e.} that these particles
are part of agglomerates showing properties similar to those of
the 17 vol\% sample.

%##########################################################################
%----------------------------------- CONCLUSIONS
\section{Conclusion}
Extensive studies of the magnetic dynamics of a nanoparticle
system containing nearly monodisperse ferromagnetic particles have
been presented. We have shown that a strongly interacting particle
system (the 17~vol\% sample) displays critical dynamics
reminiscent of that of a spin glass. Furthermore, the effect of an
Arrhenius-N{\'e}el temperature dependence of $\tau_*$ has been
explored and it has been found that dynamic scaling for this
sample prevails although with slightly reduced values of the
critical exponents $z\nu$ and $\beta$.

For weakly interacting particle systems, the dipolar interaction
will only slightly modify the relaxation compared to a
non-interacting system, and the relaxation time can be obtained by
introducing thermodynamical averages of the dipolar field in an
expression for the single-particle relaxation time, as suggested
in \cite{JonssonEPL01}. Such a theory will apply for
the most dilute sample studied here (0.06~vol\% sample) near the
blocking temperature.

For a wide range of particle concentrations (interaction
strengths), neither a pure model for weakly interacting systems
nor a model only assuming critical dynamics will correctly
describe the magnetic relaxation of an interacting particle
system. In the present work, this has been illustrated by the
5~vol\% sample, in which correlations and collective behavior are
of importance, as evidenced by non-equilibrium dynamics similar to
that exhibited by spin glasses \cite{Joensson00} and an apparent
critical slowing down in a limited temperature range. Yet, the
deviation from critical slowing down at lower temperatures and the
failure to satisfy other signatures of critical dynamics show that
the slowing down of the magnetic dynamics in this sample differs
profoundly from that in spin glasses. The range of concentrations
for which such a complex behaviour occurs becomes wider with
increasing width of the particle size distribution, as can be
evidenced by comparing to the system studied in \cite{Jonsson98}. 
For the sample used in that study, the
relative dipole interaction strength ($=M_s^2\phi/K$, where $\phi$
is the volume fraction of particles) is comparable to that of the
17 vol\% sample studied here, but that sample has a much wider
energy barrier distribution and does not exhibit critical dynamics
although it shows collective behaviour.

\ack
This work was financially supported by The
Swedish Natural Science Research Council (NFR). We would like to
thank Christian Bender Koch for the atomic absorption spectroscopy
measurements.

%**************************************************************************

%**************************************************************************

%**************************************************************************

%**************************************************************************

%**************************************************************************

%**************************************************************************

%**************************************************************************

%**************************************************************************

%**************************************************************************

\end{document}